\documentclass[12pt,superscriptaddress]{article}
\usepackage{palatino,cite,epsfig,amsmath,amssymb,graphicx}

\allowdisplaybreaks

\newcommand{\be}{\begin{equation}}
\newcommand{\ee}{\end{equation}}
\newcommand{\bea}{\begin{eqnarray}}
\newcommand{\beas}{\begin{eqnarray*}}
\newcommand{\eea}{\end{eqnarray}}
\newcommand{\eeas}{\end{eqnarray*}} 
\newcommand{\ba}{\begin{array}}
\newcommand{\ea}{\end{array}}
\newcommand{\bi}{\begin{itemize}}
\newcommand{\ei}{\end{itemize}}
\newcommand{\ben}{\begin{enumerate}}
\newcommand{\een}{\end{enumerate}}

\begin{document}

\vspace{2cm}

\begin{center}

{\Large Tracing the Gauge Origin of Yukawa and Higgs
Parameters Beyond the Standard Model}

\vspace{0.4cm}
\vspace{1cm}
{\sc 
J. Lorenzo D\'{\i}az-Cruz \\
}
\vspace*{1cm}
{\sl Cuerpo Acad\'emico de Part\'iculas, Campos y Relatividad \\
FCFM- BUAP, Puebla, Pue. 72570, M\'exico
}
\end{center}

\vspace{2cm}

\begin{abstract}

We discuss possible realizations of the hypothesis that all the 
fundamental interactions of the elementary particles should be of
gauge type, including the Yukawa and Higgs ones.  
In the minimal SUSY extension of the standard model, where
the quartic Higgs couplings are ``gauged'' through the D-terms, 
it is also possible to generate radiatively the Yukawa matrices 
for the light generations, thus expressing them as functions of
gauge couplings. The program can also be applied to the SUSY LR model,
where the possibility to induce radiatively the mixing angles, 
can help to make viable the parity  solution to the strong CP problem.
The superpotential of the model still includes some non-gauge couplings, 
namely, the Yukawa constants for the third generation and the trilinear
terms $\lambda \chi_L\Phi \chi_R$ and $\lambda'\chi^c_L\Phi \chi^c_R$,
involving the Higgs bi-doublet ($\Phi$) and two pairs of doublets
($\chi_L, \chi_R$ and their conjugates). Additional progress to
relate these parameters to gauge couplings, can be made by embedding 
the LR model within a SUSY model $SU(4)_W\times U(1)_{B-L}$ in five 
dimensions, where the Higgs bi-doublet is identified as the extra 
component of the 5D gauge field.

\end{abstract}


\newpage

\section{Introduction}

The Standard Model (SM) of the strong and electroweak
interactions has met with extraordinary succes; it has 
been tested already at the level of quantum corrections
\cite{radcorrs}. 
These corrections give some hints about the nature of the
Higgs sector, pointing towards the existence 
of a relatively light Higgs boson, with a mass of the order
of the electroweak scale,  $m_{\phi_{SM}} \simeq v$ \cite{hixjenser}.
 However, the SM is plagued with aesthetic and naturalness problems, 
which have motivated the search for theories beyond the SM, 
where those problems  could be solved. In particular, one would like
to have an understanding of the SM parameters.
 These parameters could be classified as follows:
\begin{enumerate}
\item {\it Dimensionless gauge parameters}, i.e. those
associated with the gauge symmetries ($g_1, g_2, g_3$ and $\theta_{QCD}$). 
\item The {\it dimensionfull parameter} of the Higgs potential $\mu^2$,
which fixes the scale of electroweak symmetry breaking (EWSB).
\item {\it Non-gauge dimensionless parameters},  i.e. the quartic Higgs 
coupling ($\lambda$) and the Yukawa matrices ($Y_f$, $f=u,d,l,\nu$),
which are not associated with a known symmetry.
\end{enumerate}

In the light of this classification, we could ask
how many ``forces'' are included in the SM. One may think
that it only contains three (gauge) forces. However,
from the point of view of quantum field theory, the parameters 
$\lambda$ and $Y_f$ describe interactions too, 
i.e. they induce the Higgs self-coupling
and the Higgs-fermion vertices. However, these ``forces''
are not associated with a gauge symmetry, and one may wonder why does 
nature allows the presence of such arbitrary non-gauge interactions.
If the gauge principle has been so successful in describing the 
electroweak and strong interactions,  one could expect that those 
non-gauge parameters should appear related to the gauge interactions in 
the theory that will replace the SM. 

  In this paper, we attempt to 
explore possible models that could realize this {\it{radical}} form 
of the {\it{gauge principle}}, where the only 
allowed parameters should be the gauge coupling constants and 
possibly gauge-invariant mass terms. Then, all the other interactions 
that appear to be of non-gauge type, should be derived from the gauge 
ones. Some of them could still be fundamental and will appear 
related to gauge forces because of the presence of extra 
symmetries, others may not be fundamental and would be induced 
as higher order effects of the gauge fields.

\bigskip 

 In fact, one of the simplest attempts to solve the
problem of quadratic divergences in the SM, through an 
accidental cancellation \cite{veltmanqd}, implies a relationship
between the quartic Higgs coupling and the Yukawa and
gauge constants, namely:
\begin{equation}
\lambda = y^2_t -\frac{1}{8} [ 3g^2 +g'^2]
\end{equation}
Unfortunately, this relation implies a Higgs
mass $m_\phi=316$ GeV, that seems already excluded.
Neverheless, this relation makes us suspect that
$\lambda$ and $y_t$ could be of fundamental type.

In what follows  we shall discuss how to formulate a successful program 
along these ideas. 
We shall show that the construction of an extension of the SM
that fulfills this program, only requires assembling several 
theoretical ideas that have appeared in the past, and will argue that 
the models discussed in this paper that achieve some success, seem
to point towards the existence of both supersymmetry (SUSY) 
and Extra-dimensions (XD) at scales beyond $O(1)$ TeV. 
We shall assume that the quartic Higgs couplings and the 
Yukawa couplings of the third generation are fundamental, 
and they are directly related to gauge couplings at tree-level,
because of supersymmetry and extra-dimensions. 
On the other hand, the Yukawa matrices for the light generations 
will be derived from the fundamental couplings, and
expressed as functions of the gauge couplings and other
parameters of the theory (i.e. ratios of gauge invariant masses).

 The first realization of this program, to be discussed in the next 
section, appears in the minimal SUSY extension of the SM (MSSM). 
We shall elaborate on  how the presence of D-terms allows to relate 
the quartic Higgs coupling to the SM gauge coupling constants.  
It will be shown that in this model it is also possible to express the Yukawa 
couplings for the light generations in terms of gauge couplings,
using a radiative mechanism, i.e. from loop diagrams that involve 
the fermion-sfermion-gaugino couplings.
 This mechanism also allows to express the quark mixing angles in
terms of gauge interactions, which then helps to make viable the so called
parity solution to the strong CP problem, as given by some
left-right SUSY models, which is discussed in section 3. 
Finally, we shall consider embedding the model in extra-dimensions,
where additional progress can be made to express the couplings of chiral 
superfields that appear in the superpotential, as functions of gauge 
couplings. This could be acomplished,  for instance, within the 
context of a five dimensional SUSY model $SU(4)_W\times U(1)$, where 
the Higgs bi-doublet is identified as the extra component of the 
5D gauge field.


\section{Yukawa and Higgs parameters in the MSSM }

The minimal implementation of SUSY in fundamental particle physics
(MSSM) has met with mixed success. On the positive side one 
could count: the stabilization of the Higgs mass and
Radiative EWSB, the unification of the gauge coupling constants,
and the prediction of a Dark matter candidate.
Whereas the non-observation (yet) of the superpartners,
and the corresponding mechanisms of SUSY breaking and transmition, 
needed to make them heavy enough, are among its unpleasant aspects.

However, by its own virtues SUSY also solves the problem of the
Higgs self coupling (through the D-terms), a success that in our
opinion should be counted at the same level as the gauge coupling 
unification. Furthermore, SUSY also offers some new avenues to discuss 
the problem of the Yukawa couplings.

\subsection{The quartic Higgs coupling in the MSSM }

In the MSSM, the gauge   and Higgs particles are placed 
in gauge and chiral supermultiplets, respectively. 
However, because the SUSY formalism requires to have
equal bosonic and fermionic degrees of freedom in each supermultiplet, 
one needs to introduce the auxiliary fields, which can be eliminated
by the solution to the equations of motion. For the non-abelian 
case, the vector multiplet is written as,
\begin{equation}
V^a = (\lambda^a, v^a_\mu , D^a)  =(\lambda^a, v^a_\mu , 
\sum g \phi^\dagger T^a \phi )
\end{equation}
where $\lambda^a$ denotes the gaugino, the superpartner of the
gauge field $v^a_\mu$, $D^a$ is the (gauge) auxiliary field
and $T^a$ denote the group generators. 
The sum runs over all the scalar fields of the model, which
opens the window to incorporate Higgs bilinears into the 
vector multiplet. Then, the quartic Higgs couplings are naturally related
to the gauge couplings ($g,g'$); the quartic terms  in the Higgs potential 
will appear as follows:
\begin{equation}
V_4 = \frac{g^2}{4} [ (H^\dagger_u \tau^i H_u)^2+  
         (H^\dagger_d \tau^i H_d)^2] \\
   +\frac{g'^2}{4}[(H^\dagger_u H_u)^2-  (H^\dagger_d H_d)^2]
\end{equation}
However, the resulting value for the mass of the light Higgs
boson predicted by the model, $m_h\simeq m_Z$, is getting into conflict 
with current Higgs mass bounds ($m_h \geq 115$ GeV),
and something should come to the rescue. In first place,
large Radiative corrections can make $m_h\simeq 130$ GeV \cite{mssmhix},
while new gauge contributions could induce an even heavier 
Higgs mass \cite{timhix}.

The possibility to express the scalar quartic couplings as gauge constants,
is valid not only for the Higgs boson, but also for all the scalar 
superpartners (squarks and sleptons). Furthermore, this property of
SUSY is independent of the SUSY soft-breaking, and it
survives even in models where the superpartners are very heavy,
such  as in the more minimal MSSM \cite{moremssm} where they
could be of $O(10)$ TeV, or even of $O(M_{pl})$, as  
in the recently proposed split SUSY models  \cite{splitssm}.
 
The quartic couplings among squarks and Higgs bosons 
contribute to sfermion masses, and its effect could be tested by 
meassuring the mass-difference among scalars that only differ by their
gauge quantum numbers, for instance:
\begin{equation}
m^2_{\tilde{u}_L}-m^2_{\tilde{d}_L} =  \cos 2\beta m^2_W
= m^2_{\tilde{\nu}}-m^2_{\tilde{l}}
\end{equation}
These mass relations could be tested at the NLC \cite{japanNLC},
which will be able to verify the generality of the SUSY solution to the 
problem of relating the quartic scalar couplings to the gauge constants.

\subsection{Radiative Yukawa couplings}

SUSY also has the elements that may allow to express
the Yukawa parameters as functions of gauge couplings. Namely, 
there are certain types of SUSY interactions that involve two 
fermions and one scalar, the fermion-sfermion-gaugino vertices,
which are given in terms of gauge couplings and 
formally are similar to a Yukawa coupling, in the sense that
both involve two fermions and one scalar. 
Within the MSSM, these couplings can be closed into a loop
and generate the Yukawa parameters. 
Radiative corrections to fermion masses by SUSY loops, were
discussed some time ago, both to generate the full fermion
masses \cite{radferms}, and to correct the GUT mass
relations\cite{gutferms,myradferm}. More realistic models have been
discussed recently \cite{rfmjavier}. Henceforth, we
shall assume that the third generation masses appear at 
tree-level, while the  light masses and mixing angles
are the ones that could be generated radiatively. 

To describe our method to evaluate the generation of fermion masses 
through SUSY loop, we start by writing the diagonal masses
for d-type quarks in powers of the Cabibbo angle
$\lambda=0.22$, namely:
\begin{equation}
\bar{M}_d=diag(d\lambda^4,s\lambda^2,1) \cdot m_b
\end{equation}
where $d,s$ are $O(1)$ coefficients needed to factor out the b-quark
mass $m_b$. Then, assuming that the quark mass matrices are
orthogonal, i.e. the diagonalizing matrices satisfy
$V^d_R=V^{d*}_L$, the non-diagonal mass matrix for d-type
quarks can be expressed as: $M_d = V_{CKM} \bar{M}_d V^{\dagger}_{CKM}$.
Then, using Wolfenstain parametrization for the CKM matrix one gets:

\begin{equation}
M_{d}  =
\left( \begin{array}{ccc}
(d+s)\lambda^4  & s\lambda^3 & A\rho \lambda^3 \\
   s \lambda^3  & s\lambda^2 & A\lambda^2 \\
A\rho\lambda^3  & A\lambda^2 &    1
\end{array}\right) m_b, \qquad
\end{equation}

One can then determine whether the  entries in $M_d$ can be induced as a 
SUSY loop effect.
For the quark sector it is enough to consider the gluino-squark loop,
which gives:
\begin{equation}
M^{rad}_d=  \frac{2 \alpha_s}{3\pi}  
( \frac{m_{\tilde{g}} M^2_{LR} I(x) }{m^2_{\tilde{q}}} )
\end{equation}
where the loop integral $I(x)$ depends on the ratio of squark and 
gluino masses, 
$m^2_{\tilde{q}}/m^2_{\tilde{g}}$, and is typically of order 0.5. 
 For d-type squarks and sleptons:
\begin{equation}
M^2_{LR}= v[ A_f \cos\beta- \mu Y_f \sin\beta]
\end{equation}

Given a typical size for SUSY parameters, namely 
$m_{\tilde{q}}=m_{\tilde{g}}= A_f = O(500) $ GeV, one gets a
natural size for the induced quark mass of the order
$v\times 10^{-2}\simeq 1$ GeV or less, while for leptons
the corresponding value is about 0.1 GeV, which seems suitable 
to generate masses for the first and second families.

Then, to reproduce the mass hierarchy that appears in equation (6), 
we shall assume that the trilinear A terms have precisely that hierarchy, 
and therefore the LR mass term too. Thus, one can write:
\begin{equation}
[M^2_{LR}]_{ij}= \gamma_{ij} \lambda^{n_{ij}} m^2_{\tilde{q}}
\end{equation}
where $\gamma_{ij}$ are parameters of order one and the integers $n_{ij}$
are choosen to have the same pattern as in equation (6), 
e.g. $n_{11}=4, n_{12}=3$, 
etc. Then, we have:

\begin{equation}
(M^{rad}_d)_{ij}= \frac{2 \alpha_s I(x)m_{\tilde{g}} }{3\pi} \gamma_{ij} 
\lambda^{n_{ij}}
\end{equation}

 In order to get the correct values for a particular mass matrix
entry,  we have just to verify that the  condition $M_d=M^{rad}_d$ 
can be satisfied. This will amount to find solutions to the system of 
equations for the parameters $d,s,A,\rho$  written in terms of
the spectrum of SUSY particles and the parameters $\gamma_{ij}$.
Indeed, one can see that
to get a correct d-quark mass we need to satisfy:
$\gamma_{11} m_{\tilde{g}} \simeq  6\pi (s+d)m_b/\alpha_s$, which 
only requires 
$m_{\tilde{g}}=O(1)$ TeV for $s,d,\gamma_{11}\simeq 1$. 
For the strange quark one can also get a correct mass
with reasonable values of parameters, and similar results hold
for the u-, c-quarks and charged leptons \cite{rfmjavier}. However, 
one would need a very large gluino mass to generate the b-quark mass,
which we did not assume possible anyway.

 A complete program  to generate all the light fermion masses
and the mixing angles, requires that the elements of $M^2_{LR}$, 
and the SUSY parameters needed to generate the correct textures, 
should not be in conflict with current FCNC bounds \cite{masiero}.
 For d- and s-quarks, since only the diagonal elements of $M^2_{LR}$ 
are needed, there are no problems with SUSY-induced FCNC. 
Furthermore, even when one generates the Cabibbo angle through the 12 element
of $M^2_{LR}$, there are no problems neither, because the bounds
are of the same order ($\lambda^3$) that we assumed for the
12 entry of $M^2_{LR}$.
In addition, one also needs a SUSY breaking scheme that generates the 
correct patern of soft-breaking terms. In fact, one would still need to
find out how to relate the soft-breaking terms to gauge couplings,
but to discuss this problem and its possible solutions we need to address
physics at much higher scales, of $O(100)$ TeV in gauge mediated models
 or even higher, $O(M_{pl})$ in SUGRA models \cite{softsusy}
 Recently, radiative fermion masses were studied within a model with U(2) 
flavor symmetry \cite{rfmjavier}, which also discussed how to get the
soft SUSY breaking pattern. 
Other effects of these SUSY loops for top and Higgs decays are discussed 
in \cite{ourtophix}.

 Finally, to have complete success one would also like to relate the
third generation Yukawa couplings to gauge couplings. For the top quark, 
its Yukawa constant is so large that it seems difficult to conceive  
models where it could come from loop effects, since this will require 
$\gamma_{33} m_{\tilde{g}}= O(100)$ TeV.
 However, the fact that the top Yukawa is actually of the order 
of the SM gauge couplings, opens the window to other ideas, such
as gauge-Higgs unification in extra dimensions, which will be explored in 
the final section, before that we shall dwell into the strong CP problem.


\section{Radiative Yukawas and the parity solution to the strong CP problem }

 Despite the success of the MSSM, it says little about other open issues of 
the SM, such as the flavor and CP problems. The MSSM also bring other problems 
in its own, such as the mu problem or the lepton (L) and Baryon (B) number
non-conservation. Moreover, the non-observation of the light Higgs boson, 
and the superpartners, have re-introduced  fine-tunning problems in the 
MSSM \cite{kaneftun}, which have motivated the search for alternatives or 
extensions of the MSSM.

One of these major problems is associated with the CP-violating parameter
$\bar{\theta}$ that characterize the non-trivial vacuum structure of the
QCD lagrangian. The term $\bar{\theta} G_{\mu\nu} {\tilde{G}}^{\mu\nu}$ 
contributes to the neutron
electric dipole moment ($d_n \simeq 5\times 10^{-16}\bar{\theta}$ e-cm).
In order to avoid conflict with current bounds $d_n \leq 6.3 \times 10^{-26}$,
one has to assume that $\bar{\theta} \leq 10^{-10}$, which is considered
another hierarchy problem \cite{pecceirev}.

 Within the general MSSM, there are a plethora of new phases associated with
the gaugino and sfermion soft-breaking lagrangian. In particular, the phase
of the gluino mass contributes to the $\bar{\theta}$ parameter, and it has to
be highly suppressed too; this is called the SUSY CP problem. Thus,
in the MSSM one needs to explain the size of the effective parameter 
$\bar{\theta}_{ef}$, which includes the following
contributions:
\begin{equation}
\bar{\theta}_{eff} = \bar{\theta}_{QCD}+ \bar{\theta}_{EW} + 
\bar{\theta}_{\tilde{g}}
\end{equation}
$\bar{\theta}_{QCD}$ arises from QCD instanton effects, 
$\bar{\theta}_{EW}=argDet[M_u M_d]$ is associated with chiral symmetry 
breaking in the electroweak (EW) sector
and $\bar{\theta}_{\tilde{g}}=-3arg  (m_{\tilde{g}})$ comes from
SUSY breaking.

A beautiful solution to this problem can be found within the left-Right
(LR) symmetric extension of the MSSM \cite{lrsusyo}. 
In this case the presence of the parity symmetry (P) makes the
parameter $\bar{\theta}_{QCD}=0$. In addition, the model requires the quark 
mass matrices to be hermitian, while the gluino mass is real, and these 
conditions are sufficient to have $\bar{\theta}_{eff}=0$ at tree-level
\cite{lrsusya}. Furthermore, 
it was shown in ref. \cite{lrsusya}, that this solution survives at
higher orders, since only a safely small $\theta_{eff}$
is generated at one-loop level.

The SUSY LR model has other virtues, such as making more natural
the assumption of $R-$parity conservation in the MSSM, thus protecting
the proton from decaying too fast. It also relates the hypercharges of
quarks and leptons with the gauged $B-L$ number.
The particle content of the model, including the Higgs sector, 
is shown in Table 1, which dispalys 
their quantum numbers under the gauge group
$SU(2)_L\times SU(2)_R \times U(1)_{B-L}\times SU(3)_c$.
The breaking of the gauge symmetry can occur in several steps.
For instance, one can break first: 
$SU(2)_R\times U(1)_{B-L} \to U(1)_Y$, using the
Higgs doublets $\chi_R, {\chi_R}^c$. 
Breaking of the EW symmetry:  $SU(2)_L\times U(1)_Y \to U(1)_{QED}$, 
can be done with another pair of Higgs doublets $\chi_L$ and $\chi^c_L$. 
Furthermore, to generate masses for quarks and leptons one needs 
to include the Higgs bi-doublet ($\Phi$).

\bigskip

\begin{center}
\begin{tabular}{|c|c|c|c|c|}
\hline

         &  $SU(2)_L$ & $SU(2)_R$ &  $U(1)_{B-L}$ &  $SU(3)_c$  \\
\hline
  $Q_L$    & 2    &   1       &  $\frac{1}{3}$   &    3  \\
\hline
  $Q_R$    & 1    &   3       &  $-\frac{1}{3}$   &    3  \\
\hline
  $L_L$    & 2    &   1       &  -1   &    1  \\
\hline
  $L_R$    & 1    &   2       &  1   &    1  \\
\hline
 $\Phi$    & 2    &   2       &  0   &    1  \\
\hline
  $\chi_L$    & 2    &   1       &  0  &    1  \\
\hline
  $\chi_R$    & 1    &   2       &  0   &    1  \\

\hline

\end{tabular}
\end{center}

However, with a single Higgs bi-doublet the model is unable to generate
quark mixing, while neutrinos remmain massless. Several solution
based on the inclusion of additional Higgs multiplets have been
proposed in the literature; for instance adding another bi-doublet
can allow the generation of quark mixing \cite{lrsusya}, while Higgs triplets
are used to generate neutrino masses \cite{goranetal}.

 In this paper we want to keep the minimum number of Higgs multiplets,
which should make easier our hope to relate their interactions
to gauge couplings. Thus we shall stick to the minimal model,  
which only includes the Higgs doublets and a single Higgs bi-doublet.
This choice assures that the conditions needed to keep the solution
to the strong CP problem are respected \cite{lrsusya}. The 
superpotential of the model is then:
\begin{equation}
W= m \chi_L {\chi}^c_L+m' \chi_R {\chi}^c_R + 
   \lambda \chi_L \Phi \chi_R +  \lambda' \chi^c_L \Phi \chi^c_R
    +Y_q Q_L \Phi Q_R + Y_l L_L \Phi L_R
\end{equation}
Although $W$ still allows for the presence of the diagonal Yukawa 
matrices, their number is substantialy reduced because the
model implies up-down unification ($Y_d=Y_u$). In fact we only need to 
assume the presence of the Yukawa constants for the third generation, 
as the light ones could be induced radiatively too. The couplings 
$\lambda, \lambda'$ are also arbitrary, i.e. of non-gauge type, 
however, in the next section we shall discuss how to extend the LR model 
in such a way that they are related to gauge couplings too.

Then, we can use the SUSY loops to generate quark mixing and
neutrino masses.
 Although SUSY loops based on the soft-terms $M^2_{LL}$,
have been employed previously in the literature for the SUSY LR model 
\cite{babumohap}, we believe that our formulation, which is based on the soft 
terms $M^2_{LR}$,  allows to write the conditions needed to find 
a successful model in a simpler and more concise form.
 In fact, the formulae presented in the previous section are
quite general, such that one can get the conditions
needed to generate quark mixing.
 For instance, to generate the cabibbo angle, one needs to have
$M^d_{12}= s\lambda^3=M^d_{rad}$, which then means that the
SUSY parameters must satisfy: 
$\gamma_{12} m_{\tilde{g}}= 3\pi s m_b/\alpha_s$, and again this only
requires $m_{\tilde{g}}=O(1)$ TeV. Further, having 
$M^2_{LR}\simeq \lambda^3 {\tilde{m}}^2$,
is not in conflict with current bounds on FCNC.

On the other hand, to discuss neutrino masses, one 
could use the radiative mechanism through the $W_L-W_R$ mixing 
\cite{changradnus},
or alternatively, we can follow the discussion of refs. \cite{nusbkly}, 
where it was shown how to induce the Yukawa couplings for the 
left-handed (LH) neutrinos, using the sneutrino-neutralino loop.
Assuming the presence of a trilinear term ($A_\nu$) that
induces a mixing among the LH and RH sneutrinos, and a mixing angle 
($\delta$) in the RH sneutrino sector, one obtains the following
expression for the light neutrino masses,
\begin{equation}
m_\nu \simeq \frac{g^2}{384 \pi^2} \frac{A^2_\nu v^2 \delta^2}{\tilde{m}_s} 
\end{equation}
Taking $m_{\tilde{\nu}_L} \simeq m_{\tilde{n}} \simeq m_{\tilde{B}} $
allows to generate neutrino masses of order $0.1-1$ eV, which is in the
correct range to explain atmospheric neutrino data; similar loop effects can
also explain the solar neutrino data. A complete model of neutrino masses
along these lines will be presented elsewhere \cite{futurework}.

\section{Yukawa and Higgs self-couplings from extra-dimensions.}

Although a significant reduction of couplings has  been achieved
in the LR SUSY model discussed in the previous section, we still need to
worry about the interaction between the Higgs bi-doublet and the doublets 
that appear in the superpotential, equation (12), and the large Yukawa
coupling for the third generation. 
 We shall show in this section that by embedding
the model in extra-dimensions, it is possible to achieve further 
progress to fullfill our program.

Theories with extra dimensions have received much attention
recently, mainly because of the possibility they offer to
address the problems of the SM from a new geometrical perspective.
 These range from a new aproach to the hierarchy problem \cite{ADD,RanSundrum}
up to a possible explanation of flavor hierarchies in terms of
field localization along the extra dimensions \cite{nimashmaltz}.
Model with extra dimensions have been applied to neutrino physics 
\cite{abdeletal}, GUT models \cite{hallnomu} and Higgs phenomenology 
\cite{myhixXD}, among many others.

 An interesting scenario within the extra-dimensional (XD) approach, 
consists in identifying the Higgs boson as a component of an XD 
gauge field \cite{XDGHix}. Promising models could be constructed in
five and six dimensions, with or without SUSY \cite{ABQuiros,CGMurayama}.
 Unification of Higgs and matter with the gauge multiplets has also
been discussed \cite{gauhixyuku}. 

 To illustrate the idea of symmetry breaking through orbifolds, we shall 
consider a gauge theory in 5D with a gauge group G, which is compactified on a
$S^1/Z_2$ orbifold. The 5D gauge bosons are $A_M= T^A A^A_M$  
[$M=(\mu,5)$]. The full gauge symmetry can be broken by the orbifold 
boundary conditions (O.B.C.):
\begin{eqnarray}
A_\mu(x_\mu,y) \to A_\mu(x_\mu,-y)&=& +P A_\mu(x_\mu,y)P^{-1}, \\
A_5(x_\mu,y) \to A_5(x_\mu,-y)  &=& - P A_5(x_\mu,y)P^{-1},
\end{eqnarray}
$P$ acts on gauge space as an ``inner automorphism'',
such that the gauge symmetry is broken: $G \to H$.
Thus, O.B.C. split the group generators into two sets,
$T^A=\{T^a, T^k\}$, $T^A \, \epsilon  G$, $T^a \, \epsilon H$. Since $A^a_\mu$ 
has even $Z_2-$ parity, it has zero modes in the spectrum.
 On the other hand, $A^k_\mu$ has odd $Z_2-$ parity,
and does not have zero modes in the spectrum.
 Furthermore, $A^a_5$  (odd-odd) has zero modes, 
and its v.e.v. can break the symmetry further, 
nmaley $H \to H'$.

Within the context of $N=1$ SUSY models in 5 dimensions, the quartic
Higgs couplings are given in terms of gauge constants through the D-terms, 
while the Yukawa interactions could become arise from the covariant derivative, 
and therefore can be expressed in terms of gauge constants too.
In fact, it is worth mentioning that 5D $N=1$ SUSY is equivalent to $N=2$ SUSY,
from the 4D perspective, and a pure $N=2$ SUSY theory only has one gauge 
coupling constant, although its superpotential includes Yukawa-type 
interactions.

The SUSY LR model of the previous section could be embedded into
a 5D models,  One could consider, for instance, a SUSY model
with bulk gauge symmetry $SU(4)_W\times U(1)_{B-L}$ \cite{gauhixyuku}. 
Appropriate orbifold boundary conditions can  break the symmetry to
$SU(2)_L\times SU(2)_R\times U(1)_{B-L}$, and the extra-components
of the gauge fields include a Higgs bi-doublet $\Phi$, with the
right quantum numbers to induce diagonal Yukawa couplings.
The interaction of the Higgs bidoublet with matter is given by
the 5D SUSY lagrangian, which includes terms like,
\begin{equation}
{\cal{L}}_5  = \frac{1}{\sqrt{2}} \int d^2\theta \Psi_L (\partial_5-\Phi) \Psi_R 
\end{equation} 
where $\Psi_{L,R}$ denote the chiral matter supermultiplets 
(quarks and leptons).

 Furthermore, in order to complete the Higgs sector of the SUSY LR model,
we shall assume the presence of additional scalars, namely  the
extra pairs of Higgs doublets ($\chi_L, \chi_R$ and their conjugates)
needed to break the LR and EW symmetries. However, the couplings of these
fields to the Higgs bi-doublet will also arise from  terms similar to those 
appaearing in equation (16). 
Therefore the couplings will satisfy $\lambda= \lambda'$, and will only depend 
on the 5D gauge coupling constant ($g_5$) and the compactification 
radius ($R$). 
 This result will have implications for the Higgs spectrum that will
be explored in the future \cite{futurework}. Thus, by combining 
SUSY and extra-dimensions we find further progress for our program to relate
all the couplings in terms of gauge couplings.


\section{Comments and conclusions}

In this paper we have proposed a program  to achieve a gauge 
description for all the interactions of the elementary particles, 
including the Yukawa and  Higgs ones.  
In the minimal SUSY extension of the standard model, we have 
reviewed how the D-terms allow to express the quartic Higgs couplings 
in terms of gauge constants. Furthermore, we also argued that
the Yukawa matrices for the light generations can be expressed  
in terms of gauge constants by generating them 
radiatively (through gaugino-sfermion loops). The SUSY radiative 
mechanism can in turn help to make viable the parity  solution 
to the strong CP problem. This is realized for instance in the
SUSY left-right model with a Higgs sector that includes only a
single Higgs bi-doublet ($\Phi$) and two pairs of doublets
($\chi_L, \chi_R$ and their conjugates). The superpotential
of the model inludes two non-gauge couplings, for the trilinear
terms $\chi_L\Phi \chi_R$ and $\chi^c_L\Phi \chi^c_R$, which
can be ``gauged'' by embedding the model in extra-dimensions; for instance 
by working within the context of a SUSY model $SU(4)_W\times U(1)_{L-R}$  
in five dimensions, where the Higgs bi-doublet is identified 
as the extra component of the 5D gauge field.

Thus, one needs both SUSY and extra-dimensions, in order to
make further progress to express all the 
Yukawa and Higgs parameters in terms of gauge 
couplings. However, the origin of the SUSY soft breaking terms
remmains as an open problem. This issue may have to wait for
some experimental input or further progress on the theoretical
side. In this regard, having found some models that
show at least some partial progress in expressing most of its
couplings as functions of gauge couplings, can be considered
as a possible footprint of String Theory at low energies.
String Theory would fullfill completely this program, from a
top-bottom approach,  since it starts with a single 
coupling at the Planck scale and claims that all other 
couplings should be  derived quantities. 

{\bf{Acknowledgements}} I would like to thank G. Senjanovic for the invitation
to ICTP in the summer (2004), and A. Aranda, J. Ferrandis,
P.F. Perez and A. Rosado for discussions and comments.
This work was supported by CONACYT and SNI (Mexico).


\end{document}